\documentclass[11pt]{article}
\usepackage{graphicx} % Required for inserting images
\usepackage{amsfonts,amssymb,amsmath,amsthm,cite}
\usepackage{slashed}
\usepackage{hyperref}
\usepackage{orcidlink}
\evensidemargin=\oddsidemargin
\usepackage{xcolor} \definecolor{darkgreen}{rgb}{0,.5,0}

\newcommand{\tr}{\mathrm{tr}\,}
\newcommand{\ii}{\mathrm{i}}
\newcommand{\MM}{\mathcal{M}}
\newcommand{\dd}{\mathrm{d}}
\newcommand{\nn}{\mathbf{n}}

\newcommand{\Hilb}{\mathcal{H}}

\title{Heat-Kernel approach to the Atiyah--Singer Index of Non-Hermitian Dirac Operators}
\author{Jo\~ao Pedro Breveglieri da Silva \orcidlink{0000-0003-4672-4090}\thanks{jpbreveglieri@gmail.com} and Dmitri Vassilevich\,\orcidlink{0000-0002-2894-4121}\thanks{dvassil@gmail.com} \\ Center of Mathematics, Computation and Cognition, Universidade Federal do ABC\\
 09210-580, Santo Andr\'e, SP, Brazil} 
\date{\today}

\begin{document}
\maketitle
\begin{abstract}
    If an operator $H$ anticommutes with a chirality operator $\Gamma_*$ such that $\Gamma_*^2=1$, the null space of $H$ can be decomposed in a direct sum of two spaces having positive and negative chiralities, respectively. When both spaces are finite dimensional, one can define an index, $\mathrm{Ind}(\Gamma_*,H)$, as the difference of dimensions of these two spaces. The key issue is whether $\mathrm{Ind}(\Gamma_*,H)$ is topologically protected, i.e., whether it remains constant under smooth variations of the parameters and background fields entering $H$. For Hermitian Dirac operators, topological protection of the index is guaranteed by the Atiyah--Singer theorem. In this paper, by using the heat kernel methods, we show that  $\mathrm{Ind}(\Gamma_*,H)$ is topologically protected also for non-hermitian operators $H$ as long as they are diagonalizable and satisfy some ellipticity conditions. The index is given by a heat kernel coefficient which is an integral of a local expression depending on background fields.
\end{abstract}

\section{Introduction}\label{sec:intro}
It is a text-book Quantum Mechanics that non-Hermitian Hamiltonians describe open quantum systems and thus are of interest in physics. 
Among non-Hermitian Hamiltonians the $\mathcal{PT}$ symmetric ones play a special role \cite{Bender:1998ke,Bender:2007nj}. They have a real spectrum though the eigenfunctions fail to form an orthonormal basis in the Hilbert space. In condensed matter physics, an increase of the interest to non-Hermitian systems is related, at least partially, to the non-Hermitian skin effect \cite{PhysRevB.97.121401,PhysRevB.99.201103,Liang:2022mux,Rijia:2023}.  For a recent review on non-Hermitian physics we refer the reader to \cite{Ashida:2020dkc}.

The Atiyah--Singer index theorem\footnote{This theorem was established in a series of papers starting with \cite{Atiyah:1963zz} along almost a decade. For a history survey, see \cite{bleeckerindex}.} is arguably one of the most spectacular achievements of the 20th century Mathematics. In applications of this theorem to quantum physics, one considers a Hermitian Dirac operator $H$ which anticommutes with a chirality operator $\Gamma_*$ satisfying $\Gamma_*^2=1$. The zero modes of $H$ can be separated in two sets, one with modes having a positive chirality (belonging to the eigenspace of $\Gamma_*$ with eigenvalue $+1$), the other - with a negative chirality (belonging to the space where $\Gamma_*=-1$). The difference between the dimensions of the positive chirality null space and of the negative chirality null space is called the index. The Atiyah--Singer theorem tells us that the index is a topological invariant and even provides an expression for this invariant.  This theorem has been very important for the understanding of quantum anomalies and topological structure in Quantum Field Theory. With the advancement of Dirac materials \cite{CastroNeto:2007fxn,Armitage:2017cjs}, the Atiyah--Singer theorem has also found applications in condensed matter physics where it has been useful to control the number of topologically protected states. For a proof of the index theorem by quantum mechanical methods we refer the reader to \cite{Zhou:2017llc}.

Note, that the Atiyah--Singer theorem not only states topological invariance of the index, but also provides its geometric interpretation. This latter task is beyond the scope of present work.

The analysis of topologically protected states was one of the motivations for the recent work \cite{Roy:2025llo} which addressed implications of the Index Theorem for the spectrum of non-Hermitian Dirac Hamiltonians. In \cite{Roy:2025llo}, the author analyzed non-Hermitian deformations of several particular Dirac Hamiltonians and, by computing the zero modes, demonstrated that their number is preserved under deformations. Note, that in the examples considered in \cite{Roy:2025llo}, all zero modes were of one chirality only so that the Index Theorem guaranteed topological protection of the complete number of zero eigenstates. 
The paper \cite{Roy:2025llo} motivated us to consider a generic non-Hermitian Dirac Hamiltonian and adapt the heat kernel proof of the Index Theorem (see \cite{Fursaev:2011zz} for a physics friendly exposition) to the non-Hermitian case. We relate the index to a heat kernel coefficient of $H^2$. This coefficient is, on one hand, a smooth function of background fields. On the other hand, it is an integer. Thus, the index cannot change under smooth variations of background fields. I.e., it is topologically protected. The proof goes through under two assumptions. First, $H$ should have a complete spectrum, i.e., it has to be diagonalizable. The eigenfunctions do not need to be orthogonal and the eigenvalues do not need to be real. Pseudo-Hermitian operators used in Quantum Mechanics usually satisfy this assumption. Second, the operator should be strongly elliptic. Roughly speaking, this means that the absolute values of the imaginary parts of eigenvalues of the principal symbol of the operator should be smaller than the absolute values of their real parts.

We would like to mention an alternative approach to non-Hermitian index \cite{Gong:2018uyu} which does not rely on the heat kernel expansion but uses the mathematical machinery of $K$ theory. Topological invariants for non-Hermitian lattice fermions interacting with a gauge field were studied in \cite{Guo:2021sjp,Ma:2024zbl}.

This paper is organized as follows. In the next Section, we derive a heat-kernel expression for the Atiyah--Singer index of non-Hermitian Dirac Hamiltonians satisfying the restrictions mentioned above and show that the index is topologically portected. Section \ref{sec:ex} contains examples. Concluding remarks are given in Section \ref{sec:conc}. Useful heat kernel relations are collected in Appendix \ref{sec:hk}. An example of a matrix Hamiltonian is considered in Appendix \ref{app:findim}.

\section{General scheme}\label{sec:gen}

We start with main definitions of the index theory \cite{bleeckerindex}. Consider an operator $H$ acting on a Hilbert space $\Hilb$ with an inner product $\langle\ ,\ \rangle$. In this Section, we suppose for simplicity that $\Hilb$ is separable. That is, $\Hilb$ has a countable basis. Mainly, we will deal with the case when $\Hilb$ is the space of square integrable sections (spinors) of a vector bundle $\mathcal{S}$ over a compact Riemannian manifold $\MM$, $\dim \MM =n$, with or without boundaries. If $H$ is Fredholm, i.e., if both $H$ and its conjugate have finite numbers of zero modes, one defines the Fredholm index of $H$ as
\begin{equation}
    \mathrm{Ind}(H)=\dim \ker H - \dim \ker H^\dag .\label{defindex}
\end{equation}
Let us suppose that there is a chirality operator $\Gamma_*$ satisfying $\Gamma_*^2=1$. Then, $\Hilb$ can be decomposed in a direct sum, $\Hilb=\Hilb_+ \oplus \Hilb_-$, where $\Hilb_\pm = \frac{1}{2}(1\pm \Gamma_*)\Hilb$, so that $\Gamma_*\Hilb_\pm =\pm \Hilb_\pm$. If $\Gamma_*$ anticommutes with $H$, $\{ \Gamma_*,H \}=0$, then $H$ maps $\Hilb_\pm$ to $\Hilb_\mp$. This means that $H$ can be written in block antidiagonal form,
\begin{equation}
    H=\begin{pmatrix}
        0 & H_- \\ H_+ & 0 
    \end{pmatrix}.\label{blockH}
\end{equation}
Later on we will also need
\begin{equation}
    H^2=\begin{pmatrix}
        H_-H_+ & 0 \\ 0 & H_+H_- 
    \end{pmatrix}.
\end{equation}

A $\Gamma_*$ index of $H$ is defined as
\begin{equation}
    \mathrm{Ind}(\Gamma_*,H)=\dim \ker H_+ - \dim \ker H_- .\label{GstarIndex}
\end{equation}
In other words, $\mathrm{Ind}(\Gamma_*,H)$ is the difference between the number of zero modes of $H$ of positive chirality and the number of zero modes of negative chirality. This index is usually called the Atiyah--Singer index.

If $H$ is Hermitian, $H_-=H_+^\dag$ and $\mathrm{Ind}(\Gamma_*,H)=\mathrm{Ind}(H_+)$, so that the Fredholm index (\ref{defindex}) and the Atiyah--Singer index (\ref{GstarIndex}) are interrelated. 

An important role will be played by pseudo-Hermitian operators. These are the operators for which exists an invertible operator $O$ such that
\begin{equation}
    \widehat{H}=O^{-1}\, H\, O \label{pseudoherm}
\end{equation}
is Hermitian. Basically, $O$ maps the vectors from an orthonormal basis of $\Hilb$ to eigenvectors of $H$, see \cite{Mostafazadeh:2001nr} for details\footnote{The paper \cite{Mostafazadeh:2001nr} uses a little different nomenclature. We will use the name pseudo-Hermitian for the operators $H$ satisfying (\ref{pseudoherm}) since this class will play more important role in our analysis.}. As a usual precaution, we assume that both $O$ and $O^{-1}$ are bounded \cite{Krejcirik:2014kaa}. Otherwise, the eigenvectors may not form a reasonable basis allowing to compute traces. 

We will actually work with another class of operators. An operator $H$ is \emph{diagonalizable} if and only if it has a complete basis\footnote{Here we mean a Schauder basis which is a countable basis such that each element of $\Hilb$ can be represented uniquely as an infinite series over the element of this basis convergent in $\Hilb$. This has to be contrasted to the notion Hamel basis which is frequently used in Linear Algebra and allows finite sums only. } of eigenvectors. These eigenvectors do not need to be orthogonal. Neither do the eigenvalues need to be real. Pseudo-Hermitian Hamiltonians considered in Quantum Mechanics are usually diagonalizable \cite{Mostafazadeh:2001nr,Bender:1998ke,Bender:2007nj}.

One can introduce another chirality operator
\begin{equation}
    \widehat{\Gamma}_*=O^{-1}\Gamma_*\, O \label{Gammahat}
\end{equation}
with the properties
\begin{equation}
    \widehat{\Gamma}_*^2=1,\qquad \{ \widehat{\Gamma}_*,\widehat{H} \}=0. \label{hatHGam}
\end{equation}
Obviously, the zero modes of $H$ with positive (negative) $\Gamma_*$ chirality are in one to one correspondence with the zero modes of $\widehat{H}$ with positive (negative) $\widehat{\Gamma}_*$ chirality. Therefore,
\begin{equation}
    \mathrm{Ind}(\Gamma_*,H)=\mathrm{Ind}(\widehat{\Gamma}_*,\widehat{H}).\label{indind}
\end{equation}
We can define $\widehat{\Hilb}_\pm$ and $\widehat{H}_\pm$ simply by putting hats in the definitions above. 

The $\Gamma_*$ index of $H$ can be studied within the heat kernel approach \cite{Gilkey:1973,Atiyah:1973ad,Atiyah:1975jf}. Our exposition is closer to a simplified version \cite{Fursaev:2011zz}. First, we demonstrate two important statements about the spectra of $H$ and $H^2$. This will allow us to express the index through a heat kernel coefficient of $H^2$ and thus to show that it is topologically protected. Later on, we will formulate analytic conditions under which the heat kernel exists and admits the desired asymptotic expansion.

Clearly, if $\psi$ is a zero mode of ${H}_+$ it is also a zero mode of ${H}_-{H}_+$. For a diagonalizable operator, there is a one-to-one correspondence between the eigenmodes of $H$ and $H^2$.
In particular, positive/negative chirality zero modes of $H^2$ coincide with positive/negative chirality zero modes of $H$. Thus, the zero modes of ${H}_+$ coincide with the zero modes of ${H}_-{H}_+$. The same is true for the zero modes of ${H}_-$ and ${H}_+{H}_-$. We stress, that the diagonalizabilty of ${H}$ was essential to reach these conclusions. This statement is not true for any $H$. For a simple example of a matrix Hamiltonian see Appendix \ref{app:findim}.

The next statement refers to non-zero modes. The core calculation does not require diagonalizability. Let us suppose that $\psi$ is an eigenstate of $H_-H_+$ with an eigenvalue $\lambda$ and $H_+\psi \neq 0$. Then
\begin{equation}
    (H_+H_-) H_+\psi = H_+(H_-H_+)\psi = \lambda H_+\psi 
    \label{HpHm}
\end{equation}
i.e, $H_+\psi$ is an eigenstate of $H_+H_-$ with the same eigenvalue.

Based on these two statements, we may write
\begin{equation}
    \mathrm{Ind} ({\Gamma}_*,{H})=\mathrm{Tr}\left( \Gamma_* \, e^{-t{H}^2}\right) \label{hkInd}
\end{equation}
for $t>0$.  Indeed, since ${H}$ is diagonalizable, and so is ${H}^2$, the eigenfunctions of ${H}^2$ form a basis in $\Hilb$. For each eigenfunction with chirality $+1$ there is one eigenfunction with chirality $-1$ with the same eigenvalue. Thus, the trace over the subspace spanned by non-zero modes vanishes in (\ref{hkInd}). The trace over zero modes gives $\mathrm{Ind}(\widehat{\Gamma}_*,\widehat{H})$ which, by Eq.\ (\ref{indind}), coincides with $\mathrm{Ind}(\Gamma_*,H)$.

At this point, we have to make some assumptions about the analytic properties of $H$. Basically, we have to define a class of generalized Dirac Hamiltonians for which all the previous and subsequent steps make sense. Our choice is the family of strongly elliptic first order partial differential operators. Let us proceed with the definitions.

Since Dirac Hamiltonians are unbounded, one has to be careful with defining their domains, compositions, etc. The corresponding  mathematical procedure \cite{Gilkey:1995mj} does not rely on hermiticity. Thus, we omit details.\footnote{Roughly speaking, initially $H$ is defined on the space of smooth sections of a vector bundle over $\MM$ satisfying certain boundary condition. It can be then extended to a suitable Sobolev space (which is $\mathrm{H}^1$ in our case). This space is a natural domain for $H$. It is also dense in the space $\Hilb$ of square integrable sections. To fill the gaps, the reader should consult specialized literature as, e.g., \cite{Gilkey:1995mj}. An accessible introduction to the interplay between boundary conditions and functional spaces can be found in \cite{Shubin:2020}.} 

Any first order partial differential operator can be represented in a vicinity of any point of the interior of $\MM$ as $H=\ii h_1^j(x)\partial_i + h_0(x)$ where $h_1$ and $h_0$ are some matrix valued functions. The leading symbol of $H$ is obtained by taking the terms in $H$ containing highest-order derivatives and replacing $\partial_j$ by $-\ii k_j$.\footnote{ This procedure mimics the Fourier expansion on $\mathbb{R}^n$ in terms of plane waves $e^{-\ii kx}$ with $k$ being the wave vector. The Fourier expansion of $\MM$ is quite different.} In our case, the leading symbol reads
\begin{equation}
    p(x,k)=h_1^j(x) k_j \,.\label{leadingsymb}
\end{equation}
These local expressions can glued together to a globally defined leading symbol of $H$, see \cite{bleeckerindex}. We do not need this global construction here since the notions which we will use are essentially local.
The operator $H$ is called \emph{elliptic} if the leading symbol is non-degenerate for all $x$ and all $k\neq 0$. Ellipticity guarantees many nice properties of partial differential equations involving $H$. However, since we are going to work with $e^{-tH^2}$ we need a stronger requirement  \cite{Gilkey:1983xz}. We will call $H$ \emph{strongly elliptic} if it is elliptic and for all eigenvalues $\lambda$ of the leading symbol $p(x,k)$ with $k\neq 0$ one has $|\mathrm{Re}\, \lambda|>|\mathrm{Im}\, \lambda|$. If the leading symbol is Hermitian, the requirement of strong ellipticity is obviously redundant\footnote{In this case the notion of strong ellipticity is used to characterize boundary conditions, see remarks at the end of this Section. What we call strong ellipticity is also called in the literature ``strong ellipticity in the volume" or ``ellipticity with respect to a cone in the complex plane".}. 

We stress that the condition formulated above refers to the behavior of leading symbol inside $\MM$. If $\partial\MM=\emptyset$, this is all what we need. In the presence of boundaries, one more condition has to be imposed, see at the end of this Section. 

As an example, consider
\begin{equation}
    H=\ii z (\sigma^1\partial_z + \sigma^2\partial_2) \label{strongelexample}
\end{equation}
where $\sigma^{1,2}$ are the Pauli matrices and $z$ is a complex number. For this operator, $p(x,k)=z (\sigma^1 k_1 + \sigma^2 k_2)$. Using the properties of Pauli matrices, one can easily derive that (\ref{strongelexample}) is elliptic for $z\neq 0$, strongly elliptic for $|\mathrm{Re}\, z|> |\mathrm{Im} z|$ and Hermitian for any real $z$.

The idea behind the strong ellipticity requirement is to ensure that $p(x,k)^2$ is a quadratic form in the $k$ space with eigenvalues having positive real parts. This makes $\exp( -t p(x,k)^2)$ an integrable function of $k$ for any positive $t$. This is enough to guarantee that the heat operator $e^{-tH^2}$ is trace class for $t>0$. Moreover, for any matrix-valued function $Q$ there is an asymptotic expansion at $t\to +0$ (see Lemma 2.6 in \cite{Gilkey:1983a})
\begin{equation}
    \mathrm{Tr} \left( Q e^{-tH^2} \right) \simeq \sum_{l=0}^\infty t^{\frac{n-l}{2}} a_l(Q,H^2) ,\label{asymp}
\end{equation}
where the heat kernel coefficients $a_l$ are smooth local functionals of $Q$, parameters and fields appearing in $H^2$, and their derivatives.

Since the left hand side of (\ref{hkInd}) does not depend on $t$, we can compute the right hand side for any value of $t$. Let us do this for $t\to 0$. All divergent terms in the expansion (\ref{asymp}) with $Q=\Gamma_*$ should vanish and the index should be given by the constant term with $l=n$,
\begin{equation}
    \mathrm{Ind} ({\Gamma}_*,H)=a_n (\Gamma_*,H^2) .
    \label{indan}
\end{equation}
In this equation, the left hand side is an integer while the right hand side is a smooth functional of the field. Thus, in fact, both sides of this equation cannot depend on the fields. More precisely, the index cannot change under small smooth variations of fields and parameters. I.e., it is a topological invariant.

The result of this Section can be summarized as follows. Let $H$ be a strongly elliptic diagonalizable Dirac Hamiltonian anticommuating with a chirality matrix $\Gamma_*$. Then $\mathrm{Ind}(\Gamma_*,H)$ is a topological invariant.
 
 %Since the coefficient $a_n(\Gamma_*,H^2)$ must be an integer, any operator for which $a_n(\Gamma_*,H^2)$ is not an integer is not diagonalizable. 

We conclude this Section with comments on the case when $\MM$ has a smooth boundary $\partial\MM$. Then, one has to impose boundary conditions on $\psi$. Since $H$ is a first order operator, one has to fix one half of the components of $\psi$ which can be done by choosing a projector $\Pi_-$ and requesting the Dirichlet condition
\begin{equation}
    \Pi_-\psi\vert_{\partial\MM}=0, \label{bc}
\end{equation}
see Sec.\ \ref{sec:ex2} for an example. To be able to define chiral eigenmodes one has to request\footnote{On a side note, the possibility to define chiral transformations and the chiral anomaly imposes a different restriction on the boundary conditions, see \cite{Erdmenger:2024zty}. }
\begin{equation}
    [\Gamma_*,\Pi_-]=0. \label{GammaPi}
\end{equation}
The existence of heat kernel imposes some additional strong ellipticity conditions on the boundary projector. Here, we do not go into details of these conditions and refer the reader to \cite{Gilkey:1973}. An example on non-local projector which satisfies all restrictions is given by the Atiyah--Patodi--Singer boundary conditions \cite{Atiyah:1975jf}. An example of local strongly elleiptic boundary conditions can be found in Sec.\ \ref{sec:ex2} below. More general suitable local boundary conditions are discussed in \cite{FarajiAstaneh:2023fad}.

\section{Examples}\label{sec:ex}
\subsection{Dirac Hamiltonian on a circle}\label{sec:ex1}
Consider an operator 
\begin{equation}
    H=\begin{pmatrix}
        0 & \ii \partial_x -a \\ \ii \partial_x -b & 0
    \end{pmatrix} \label{ex1H}
\end{equation}
on unit $S^1$ with some real constants $a$ and $b$. This operator anticommutes with
\begin{equation}
    \Gamma_*=\begin{pmatrix}
        1 & 0 \\ 0 & -1 
    \end{pmatrix}\label{ex1Gam}
\end{equation}
The leading symbol (\ref{leadingsymb}) of $H$ 
\begin{equation}
    p(x,k)=\begin{pmatrix}
        0 & k \\ k & 0
    \end{pmatrix} \label{ex1p}
\end{equation}
has real eigenvalues $\pm k$ so that $H$ is strongly elliptic. To find the spectrum of $H$ it is sufficient to analyze its action on the Fourier harmonics $e^{-\ii k x}$ and thus to define the eigenvectors of the matrix
\begin{equation}
    \begin{pmatrix}
        0 & k-a \\ k-b & 0
    \end{pmatrix},\qquad k\in \mathbb{Z}.\label{ex1matrix}
\end{equation}
One can easily obtain that
\begin{enumerate}
    \item The operator $H$ has a zero mode of positive chirality iff $b\in \mathbb{Z}$ and a zero mode of negative chirality iff $a\in \mathbb{Z}$.
    \item $H$ is diagonalizable except for the cases when $a\neq b$ and at least one of parameters $a,b$ is integer. The eigenvalues of $H$ may be complex.
\end{enumerate}
Thus, if $H$ is diagonalizable $\mathrm{Ind}(\Gamma_*,H)=0$. The index is topologically protected meaning that it does not change with the changes of $a$ and $b$ as long as $H$ remains diagonalizable. 

The operator $H$ is not diagonalizable if at least one of the parameters $a$ or $b$ is an integer and $a\neq b$. Such points in the parameter space are called exceptional points, or physical exceptional points \cite{Ashida:2020dkc} if one likes to stress the difference in terminology with mathematical literature \cite{kato1966perturbation}. In the exceptional points of the model considered here $\mathrm{Ind}(\Gamma_*,H)=\pm 1$ except for the case when both $a$ and $b$ are integers.

Let us see what it looks like from the heat kernel perspective. The square of (\ref{ex1H}) reads
\begin{equation}
    H^2 =\begin{pmatrix}
        1 & 0 \\ 0 & 1
    \end{pmatrix} (\ii \partial_x -a)(\ii \partial_x -b) \label{ex1H2}
\end{equation}
This operator has equal equal spectrum of positive and negative chirality modes. It is easy to see, that $H^2$ has extra zero modes as compared to $H$ exactly when $H$ is \emph{not} diagonalizable. Besides, since we are on a closed one-dimensional manifold,
\begin{equation}
    a_1(\Gamma_*,H^2)=0 \label{ex1a1}
\end{equation}
which confirms that $\mathrm{Ind}(\Gamma_*,H^2)=0$ whenever $H$ is diagonalizable.

Equation (\ref{ex1a1}) implies that $H_+H_-$ and $H_-H_+$ alwas have equal numbers of zero modes. These are the exceptional points where $H_+H_-$ (respectively, $H_-H_+$) may have more zero modes than $H_-$ (respectively, $H_+$).

\subsection{Dirac Hamiltonian on an interval}\label{sec:ex2}
Here we take $\MM=[0,\pi]$ and use the same expression (\ref{ex1H}) for the Dirac Hamiltonian as in the previous section. By a smooth unitary gauge transformation $\psi \to e^{\ii c x}\psi$ one can shift $a$ and $b$ to $a+c$ and $b+c$. In contrast to the circle, the phase factor $e^{\ii c x}$ does not need to be periodic. Thus, $a$ and $b$ can be shifted by an arbitrary number keeping $a-b$ fixed. Without loss of generality we choose 
\begin{equation}
    b=-a \label{ex2ba}
\end{equation}
and take the boundary projector in (\ref{bc}) in the form
\begin{equation}
    \Pi_-=\frac 12 (1-\Gamma_*). \label{ex2Pi-}
\end{equation}
The operator $H$ maps $\Hilb$ to itself. Thus,
\begin{equation}
    0=\Pi_-H\psi\vert_{\partial\MM}=(\ii \partial_x +a)\Pi_+\psi\vert_{\partial\MM} \label{ex22ndbc}
\end{equation}
with
\begin{equation}
   \Pi_+=\frac 12 (1+\Gamma_*) \label{ex2Pi+} 
\end{equation}
yielding mixed boundary conditions with $\chi=\Gamma_*$, see Appendix \ref{sec:hk}. In short, the negative chirality component $\psi_-$ satisfies Dirichlet boundary conditions, while $\psi_+$ obeys Robin conditions. These boundary conditions are chiral invariant since (\ref{GammaPi}) is obviously satisfied. To check strong ellipticity at the boundary one has to analyze the spectrum of an auxiliary boundary value problem which depends only on the highest derivative terms in $H$ and in boundary operators. That is, one has to retain $\partial_x$ in (\ref{ex1H}) and (\ref{ex22ndbc}) and drop $a$-dependent terms. Therefore, strong ellipticity of our boundary value problem is not effected by the non-Hermitian deformation and follows from the analysis of \cite{Gilkey:1973}. This property is however non-generic. If a non-Hermitian deformation changes the leading parts of $H$ or of the boundary operators, such deformation may also break strong ellipticity of the corresponding boundary values problem.

For all values of $a$ there is exactly one zero mode of positive chirality for which 
\begin{equation}\psi_-=0, \qquad \psi_+\propto e^{\ii a x}. \label{ex2zeromode}
\end{equation} 
There are no zero modes of negative chirality. Thus,
\begin{equation}
    \mathrm{Ind}(\Gamma_*,H)=1 .\label{ex2Ind}
\end{equation}

Let us analyse the non-zero spectrum of $H$. If $a\notin \mathbb{Z}\backslash \{ 0 \}$, for each $k=1,2,3,\dots$ there are two eigenmodes with the eigenvalues 
\begin{equation}
    \lambda=\pm\sqrt{k^2-a^2}. \label{ex2lambda}
\end{equation}
They read
\begin{eqnarray}
    &&\psi_-=C(k) \sin (kx), \nonumber \\
    &&\psi_+=\frac{C(k)}{\lambda} (\ii k \cos (kx) - a \sin (kx) ), \label{ex2modes}
\end{eqnarray}
where $C(k)$ is a normalization constant. If $|a|>1$ some eigenvalues are imaginary. One can check\footnote{The easiest way to check this statement is compare the eigenmodes of $H$ with the eigenmodes of $H^2$ with the same boundary conditions. The latter operator is selfadjoint and thus has a complete spectrum.}, that the eigenmodes (\ref{ex2modes}) jointly with the zero mode (\ref{ex2zeromode}) form a complete set in the space of spinors subject to Dirichlet boundary conditions on $\psi_-$ and Robin boundary conditions (\ref{ex22ndbc}) on $\psi_+$. Thus, $H$ is diagonalizable if $a\notin \mathbb{Z}\backslash \{ 0 \}$. 

If $a=k'$ is a non-zero integer, the modes with $k\neq k'$ are not modified, while the mode with $k=k'$ becomes the zero mode (\ref{ex2zeromode}) for which $\psi_-=0$. Thus $\psi_-\propto \sin(k'x)$ is absent in the spectrum. Since $\sin(k'x)$ is orthogonal to $\sin(kx)$ with $|k'|\neq |k|$ the spectrum is no longer complete. For $a\in \mathbb{Z}\backslash \{ 0 \}$ the operator $H$ is not diagonalizable. These values of $a$ are exceptional points in the parameter space.

 Let us compute the heat kernel expression (\ref{hkInd}) for the index. Since we are on a one-dimensional manifold, $n=1$, the boundary integral in (\ref{a1hk}) becomes a sum over two boundary points, $x=0$ and $x=\pi$, yielding
\begin{equation}
    a_1(\Gamma_*,H^2)=1 \label{ex2a1}
\end{equation}
which does not depend on $a$ and is consistent with (\ref{ex2Ind}). Eq.\ (\ref{ex2a1}) reproduces correctly the index at exceptional points though this is a coincidence. 

\subsection{Operators on a plane}\label{sec:ex3}
The case of a noncompact manifold $\MM$ is more complicated than the compact one. This refers to the spectral properties of the operator as well as to the existence of the heat kernel. The zero modes should be understood as normalizable zero modes. The properties of the heat kernel expansion depend crucially on asymptotic behavior of the background fields. For a Laplace type operator $L$, the asymptotic expansion of the local heat kernel $\langle x| e^{-tL}|x\rangle$ always has a rather standard form and can be obtained, for example, by covariant perturbative techniques \cite{Barvinsky:1990up,Avramidi:1997jy}. The coefficients in this expansion may not be integrable on the whole $\MM$. This is intimately related to the convergence of a Duhamel type expansion for the heat operator \cite{Iochum:2012bu}. If the invariants associated with $H$ (e.g., curvatures, scalar potentials, etc) vanish sufficiently fast at the infinity so that the local heat kernel coefficients are integrable, one can use the universal local formulas to calculate the index. An example is the Aharonov--Casher calculation of the index for planar fermions in a magnetic field \cite{Aharonov:1978gb}. A mathematically rigorous treatment of this point can be found in \cite{bleeckerindex}. Note, that our conclusion of topological protection of the index does not depend on the integrability of heat kernel coefficients. For example, the Dirac Hamiltonian of fermions on a line on the background of a kink contains a non-integrable scalar potential which makes the local formulas not applicable. Nevertheless, the index of this operator is topologically protected as long as the asymptotic values of scalar field do not change \cite{Jackiw:1975fn}. An overview of the heat kernel expansion and other spectral functions in the case of non-trivial asymptotics of background fields can be found in \cite{Niemi:1984vz}.

In Ref. \cite{Roy:2025llo}, several examples of non-Hermitian index for a class of Hamiltonians suggested in \cite{Juricic:2023szr} and used later in \cite{Pino-Alarcon:2026sgu} were considered. The most elaborate example is a Hamiltonian on $\MM=\mathbb{R}^2$,
\begin{equation}
    H = (1+\alpha M) \gamma^j D_j + \Delta_j \gamma^{2+j},
\end{equation}
where $D_j = -i\partial_j -eA_j$ for $j=1, 2$, $A_j$ is a vector potential, $e$ is the electric charge, and $\alpha$ is a real constant. $\{ \gamma^\mu\}$, $\mu=1,2,3,4,$ is a set of usual Euclidean $4\times 4$ Dirac $\gamma$ matrices, $\gamma^\mu \gamma^\nu +\gamma^\nu \gamma^\mu =\delta^{\mu\nu}$. $\Delta_1$ and $\Delta_2$ are some real functions of the coordinates on $\mathbb{R}^2$. $M=\ii \gamma^1\gamma^2$ is a constant matrix which commutes with $\gamma^3$ and $\gamma^4$ and anticommutes with $\gamma^1$ and $\gamma^2$. If $\alpha=0$, the operator $H$ Hermitian. With a suitable choice of potentials  $\Delta_j$ it can be identified with a Hamiltonian of spinor particles interacting with an Abrikosov--Nielsen--Olesen vortex, see \cite{Jackiw:1981ee}.

For $\alpha\neq 0$, the operator $H$ is not Hermitian. However, we can perform a similarity transformation
\begin{equation}
    \widehat{H} = O^{-1} H O, \qquad O=e^{cM}, \qquad \alpha =\tanh (2c) \label{ex3Pseudo}
\end{equation}
yielding
\begin{equation}
    \widehat{H}=\sqrt{1-\alpha^2}\, \gamma^j D_j+ \Delta_j \gamma^{2+j}. \label{ex3Hhat}
\end{equation}
The operator $\widehat{H}$ is Hermitian as long as
\begin{equation}
    |\alpha|<1 .\label{restralpha}
\end{equation}
Therefore, with this restriction on $\alpha$, the operator $H$ is pseudo-Hermitian.

 The leading symbol $\sigma_H$ of $H$ is obtained by replacing $D_j\to -k_j$ and omitting the terms with $\Delta_j$ in $H$
 \begin{eqnarray}
     \sigma_H(x, k) =-(1+\alpha M)(\gamma^j k_j)
 \end{eqnarray}
It is easy to find that $\sigma_H$ has eigenvalues 
\begin{equation}
    \lambda_\pm = \pm \sqrt{1-\alpha^2}|k|. 
\end{equation}
However, since $O$ is not unitary, the eigenvectors of $\sigma_H$ are not orthogonal. If $|\alpha|<1$, $\lambda$ is real, and if $|\alpha| >1$, $\lambda$ is pure imaginary. If $|\alpha|<1$, $H$ is strongly elliptic. For $|\alpha|>1$ the operator $H$ is elliptic but not strongly elliptic. 

The chirality operator is the the $\gamma^5$ matrix,
\begin{equation}
    \Gamma_*=\gamma^5=\gamma^1 \gamma^2 \gamma^3 \gamma^4. \label{ex3Gammastar}
\end{equation}
This operator anticommutes with $H$. Since for $|\alpha|<1$ the operator $H$ is strongly elliptic and pseudo-Hermitian (and thus diagonalizable), $\mathrm{Ind}(\gamma^5,H)$ is topologically protected. That is, $\mathrm{Ind}(\gamma^5,H)$ does not change under smooth deformations of $A_j$ and $\Delta_j$ which do not modify the asymptotic conditions. Furthermore,
\begin{equation}
    \widehat{\Gamma}_*=O^{-1}\Gamma_* O=\gamma^5 .\label{ex3hatG}
\end{equation}
Thus 
\begin{equation}
   \mathrm{Ind}(\gamma^5,H)= \mathrm{Ind}(\gamma^5,\widehat{H}). \label{ex3IndInd}
\end{equation}
Due to this relation, one does not even need to compute $\mathrm{Ind}(\gamma^5,H)$ independently since it coincide with the index of a very well studied Hermitian Hamiltonian. 

Our conclusions are consistent with explicit calculations of the zero mode of $H$ performed in \cite{Roy:2025llo} for a particular ansatz for the background fields. We like to stress that our result regarding topological protection of the index goes far beyond this ansatz. Moreover, the relation (\ref{ex3Pseudo}) allows one to obtain the whole spectrum of $H$ from the spectrum of the Hermitian Hamiltonian $\widehat{H}$. 

For $|\alpha|>1$, zero energy non-normalizable solutions were found in \cite{Roy:2025llo}. Non-normalizable zero modes are not controlled by the Atiyah--Singer theorem since they cannot be cleanly separated from the continuous spectrum. 

\section{Conclusions}\label{sec:conc}
We have demonstrated that if a non-Hermitian Dirac Hamiltonian $H$ is diagonalizable, strongly elliptic, and if it anticommutes with a chirality operator $\Gamma_*$, the $\Gamma_*$ index of $H$ is topologically protected. We have also presented some examples confirming our general results and illustrating the power of the methods. Here, we would like to briefly outline possible directions of the future research.

The heat kernel expression for the index (\ref{indan}) is blind to the existence of exceptional points where the diagonalizability is lost. Being naively applied at these points, (\ref{indan}) gives either a correct (as in the example of section \ref{sec:ex2}) or wrong (as in the example of section \ref{sec:ex1}) result for the index. It would be interesting to develop refined methods based on the heat kernel and  spectral function techniques allowing to detect and study exceptional points of non-Hermitian systems. At the same time, since the index is not continuous at the exceptional points at least for some examples, extending our results for non-diagonalizable operators can be problematic, if possible.

An interesting though hard problem is to give a geometric interpretation to to the topological invariants describing the index and thus explore the full power of Atiyah--Singer theorem.

Since the chiral anomaly is technically very similar to the index, we will address an extension of our methods for the non-Hermitian chiral anomaly \cite{Chernodub:2019ggz,Sayyad:2021thw,Sayyad:2023kui} in the near future.

\subsection*{Acknowledgments}
We are grateful to Ren\'{e} Meyer for provoking out interest in non-Hermitian physics and many stimulating discussions, to Andrei Smilga for correspondence, and to Alex Arvanitakis for useful comments. This work was supported in parts by the S\~ao Paulo Research Foundation (FAPESP) through the grants 2021/10128-0 (DV), 2025/02766-8 (JPBS) and by the National Council for Scientific and Technological Development (CNPq), grant 304758/2022-1 (DV). 

\appendix
\section{Heat kernel expansion}\label{sec:hk}
Here we collect some basic formulas on the heat kernel expansion, see e.g. \cite{Vassilevich:2003xt}, which are needed in the main text of this article. We consider exclusively the operators of Laplace type which can be written as
\begin{equation}
    L=-(g^{ij}\nabla_i\nabla_j +E) \label{Laplaceop}
\end{equation}
where $g^{ij}$ is a Riemannian metric, $\nabla_j=\partial_j +\omega_j$ is a covariant derivative, $E$ is a matrix valued function.

Let us assume that the boundary conditions are of mixed type 
\begin{equation}
    \Pi_-\psi \vert_{\partial\MM}=0,\qquad (\nabla_\nn +S)\Pi_+\psi\vert_{\partial\MM}=0 \label{mixedbc}
\end{equation}
where $\Pi_-$ are two complementary projectors, $\Pi_\pm =\frac 12 (1\pm \chi)$ with $\chi^2=1$, $\nn$ is an inward pointing unit normal to the boundary, $S$ is a matrix valued function on the boundary. Let $\sqrt{h}$ be induced volume element on $\partial\MM$. Then,
%Let $R$ and $k$ denote the Riemann curvature scalar and the extrinsic curvature of the boundary, respectively. Then,
\begin{eqnarray}
    && a_0(Q,L)= (4\pi)^{-n/2}\int_{\MM} \dd^n x \sqrt{g}\, \tr (Q),\label{a0hk}\\
    && a_1(Q,L)= (4\pi)^{-(n-1)/2} \tfrac 14 \int_{\partial\MM} \dd^{n-1}x \sqrt{h}\, \tr (\chi Q).\label{a1hk}%\\
   % && a_2(Q,L)=(4\pi)^{-n/2}\tfrac 16 \left[ \int_{\MM} \dd^n x \sqrt{g}\, \tr ( 6 QE +QR) \right. \nonumber\\
    % &&\qquad\qquad\qquad +\left.\int_{\partial\MM} \dd^{n-1}x \sqrt{h}\, \tr (2Qk + 3\chi \nabla_\nn Q + 12 QS) \right] \label{a2hk}
\end{eqnarray}

\section{A finite dimensional example}\label{app:findim}
Here we give a simple example of a non-diagonalizable matrix Hamiltonian which violates the relation (\ref{hkInd}) between the index and the heat trace. Let
\begin{equation}
    H=\begin{pmatrix}
        0 & 1 \\ 0 & 0
    \end{pmatrix},\qquad \Gamma_*= \begin{pmatrix}
        1 & 0 \\ 0 & -1 
    \end{pmatrix}.
\end{equation}
Obviously, $H$ has a single zero mode of positive chirality. Thus, 
\begin{equation}
    \mathrm{Ind}(\Gamma_*,H)=1.
\end{equation}
Since $H^2=0$,
\begin{equation}
    \mathrm{Tr}\bigl( \Gamma_* e^{-tH^2}\bigr)=\mathrm{Tr}(\Gamma_*)=0.
\end{equation}
Thus the condition (\ref{hkInd}) is violated.

\bibliographystyle{utphys}
\bibliography{bib-MESTRADO-JPBS}

\end{document}